\documentclass[%
reprint,
 amsmath,amssymb,
pra,
floatfix,
]{revtex4-1}

\usepackage{graphicx}
\usepackage{dcolumn}
\usepackage{bm}

\usepackage{physics}
\usepackage[mathscr]{euscript}
\usepackage{caption}
\usepackage{subcaption}
\usepackage{tikz}
\usepackage[title]{appendix}
\usepackage{float}
\usepackage{hyperref}
\usepackage{lipsum, babel}

\hypersetup{
    colorlinks=true,
    linkcolor=blue,
    citecolor=blue,      
    urlcolor=blue,
}
 
\begin{document}

\title{Half unit cell shift defect induced helical states in Fe-based chalcogenide superconductors}
\author{Tamoghna Barik $^1$}
\author{Jay D. Sau $^2$}

\affiliation{Condensed Matter Theory Center $^{1,2}$, Department of Physics $^{1,2}$ and Joint Quantum Institute $^{2}$ , \\
University of Maryland, College Park, MD-20742, USA}

\begin{abstract}
Recent scanning tunneling spectroscopy along crystalline domain-walls  associated with a half unit cell shift have revealed sub-gap density of states that are expected to arise from helical Majorana modes. Such propagating Majorana modes have been proposed to exist on the surface state of topological materials similar to FeTe$_{\text{1-x}}$Se$_\text{x}$ (FTS) along line defects where the superconducting order parameter (OP) is phase shifted by $\pi$.  Here we show that such a $\pi$ shift in theOP across the half unit-cell shift domain-wall can occur in quite conventional tight-binding models of superconducting  FTS as a result of the $s_{\pm}$ pairing symmetry across $\Gamma$ and M pockets of FTS.  
The resultant inter-pocket transmission between $\Gamma$ and M pockets is found to be typically larger than the intra-pocket transmissions. We confirm these conclusions with a calculation based on the Bogoliubov-de-Gennes (BdG) formalism which shows that a $\pi$-shift across the domain-wall is favored for a large range of model parameters for FTS. We discuss parameter regimes where this mechanism might explain the STS experiments as well as propose to test this explanation by searching for evidence of large inter-pocket scattering. 
\end{abstract}

\maketitle

\section{Introduction}
Vortex Majorana bound states (VMBS) in topological superconductors (TSC) are potential candidates for qubits in fault-tolerant quantum computation scheme due to the topological protection to its bit-state along with its non-Abelian character facilitating unitary operations \cite{Kitaev2003,Sankar_RMP2008,Sankar_npj2015}. In addition to previously known platforms for VMBS \cite{VMBS_BiTe_2015,VMBS_BiTe_2016,VMBS_Li(OH)FeAs_2018} the recent discovery of the strong topological insulator (TI) phase in the iron-based superconductor, FeTe$_{\text{1-x}}$Se$_\text{x}$ (FTS) \cite{Kanigel_band_inversion,ZhongFang2015} has attracted attention as a promising candidate due to the intrinsically induced s-wave pairing on its topological surface states \cite{Morr_TSC_FTS}. In addition to multiple promising results for realization of VMBS states in this material already \cite{Dongfeietal2018,Hanaguri2019,ShikSin2018,SCZhang2016}, further theoretical and experimental efforts for understanding certain detailed aspects of the MBS, such as conductance quantization of zero bias peaks (ZBP), inhomogeneity in ZBPs, etc., are ongoing. Apart from the Majorana zero modes (MZMs) localized at the vortex core, another characteristic feature of the TSC phase arises in the form of dispersing Majorana modes along a linear defect when the TSC gap flips sign across the defect \cite{Fu_Kane_theory}. 

\textbf{Motivation:}
In a recent scanning tunneling spectroscopy (STS) \cite{Madhavan2020} experiment evidences of such exotic dispersing Majorana modes have been observed on superconducting FTS surface along a crystalline domain-wall (DW) that is associated with a shift of the lattice by half of unit cell. The observed results were conjectured to be the result of a $\pi$-shift in the superconducting phase across the DW, which has been predicted to lead to helical Majorana DW states~\cite{Fu_Kane_theory} on TI surface states. Such states play a crucial role in schemes for manipulating MZMs at the surface of TIs.  The relatively flat (i.e., energy independent) tunneling density of states (DOS) at the DW spanning the TSC gap is consistent with linearly dispersing helical modes~\cite{Madhavan2020}. Furthermore, the absence of such feature at twin DW in the non-topological sample of FeSe provides additional evidence of its topological origin in FTS\cite{Madhavan2020}. However, alternative mechanisms for the $\pi$-phase shift such as ferromagnetic order, that would break time-reversal symmetry, has been proposed to explain the observation \cite{Magnetic_DW_theory}. Therefore, a quantitative model for how such helical modes might arise without spontaneous symmetry breaking is desirable to be able to predict more quantitative features of such modes.  

In this paper, we show that introducing an half-unit-cell-shift (HUCS) domain-wall (DW) into reasonable models for FTS can generate a $\pi$- shift in the phase of the bulk SC order parameter (OP). Since the surface TSC gap is induced by the bulk OP, a DW associated $\pi$-junction in the bulk would naturally induce the same phase shift for the TSC gap on the surface.
The origin of the bulk $\pi$-shift  of the superconducting OP is based on the mixing between the electron-like ($M$) and hole-like ($\Gamma$) pockets by the DW. The $s_{\pm}$ pairing state in FTS \cite{s_pm_Hanaguri,Chen2022} implies that the OPs associated with the two pockets ($\Gamma$ and $M$) have a relative $\pi$ phase difference. Thus, strong mixing of the electronic modes between $\Gamma$ and $M$ pockets translates to coupling between the OPs of the two pockets belonging to the opposite sides of the DW via Andreev reflection. This would imply, as we will elaborate in Sec. \ref{sec:mechanism}, that a $\pi$-junction would accommodate Andreev bound states (ABS) that have lesser energy for the occupied levels compared to the  trivial junction with no phase shift. Thus, the strong inter-pocket mixing in a background of $s_{\pm}$ pairing gives rise to the energetic stability of a $\pi$-junction across the half-shifted domain-wall. We numerically address the junction stability by comparing the ABS energies from the BdG spectrum of $0$- and $\pi$-junction, calculated on an effective model of FTS monolayer.

Since the surface TSC gap is induced by the bulk pairing of subsequent layers, a $\pi $-shift in TSC gap would occur only when the shifted domain-wall continue beneath the top surface up to few layers into the bulk. The energy cost associated with such defect is indeed higher than a single layer shift. However, we would like to point out that the extra cost is negligible w.r.t. the energy cost of a single layer shift. This is because the main contribution to the energy cost is due to the two-dimensional interface between the last shifted layer and the subsequent un-shifted layer in which case the number of mismatches is proportional to the interface area whereas the extra cost for shifting each additional layer from top adds to number of mismatches that is proportional to the DW length, hence, it is thermodynamically insignificant. Thus, continuation of the half shifted DW up to few layers is effectively as likely as having just the top layer shifted. In that case the $\pi $-shift in the bulk OP of the layers beneath would induce a $\pi$-junction on the TSC surface.

\textbf{Outline:}
In Sec. \ref{sec:mechanism} we start by elaborating on the intuitive idea for the $\pi$-shift mechanism in a bulk FTS layer by considering the two limiting cases of scattering, viz. i) pure intra-pocket transmission and ii) pure inter-pocket transmission. For these two cases we contrast the analytical approximation of the ABS energies between the 0-junction and $\pi$-junction. Thus, building an intuition for the inter-pocket transmission favoring the $\pi$-junction, we move on to investigate the junction energetics quantitatively. We first construct an effective model of an FTS monolayer in Sec. \ref{sec:effective_model} based on the tight binding approach. The model parameters are tuned to obtain the Fermi pockets in reasonable agreement with the ARPES measurements. Then incorporating the HUCS defect we describe the numerical evaluation of the DW-induced scattering problem using KWANT in Sec. \ref{sec:scattering_problem}. In Sec. \ref{sec:BdG_spectrum} we describe the BdG spectrum calculation for the 0-junction and $\pi$-junction with an $s_{\pm}$ pairing Hamiltonian. By comparing the resultant energy levels the stability of the junctions (0- and $\pi$-) are compared. Finally, in Sec. \ref{sec:correlation} we combine results from the two calculations (Sec.\ref{sec:scattering_problem} and \ref{sec:BdG_spectrum}) to quantitatively correlate stronger inter-pocket transmission amplitudes to the $\pi$-junction stability.

\section{Qualitative origin of $\pi$-junction\label{sec:mechanism}}
\begin{figure}[h]
	\centering
    \includegraphics[width=.8 \columnwidth]{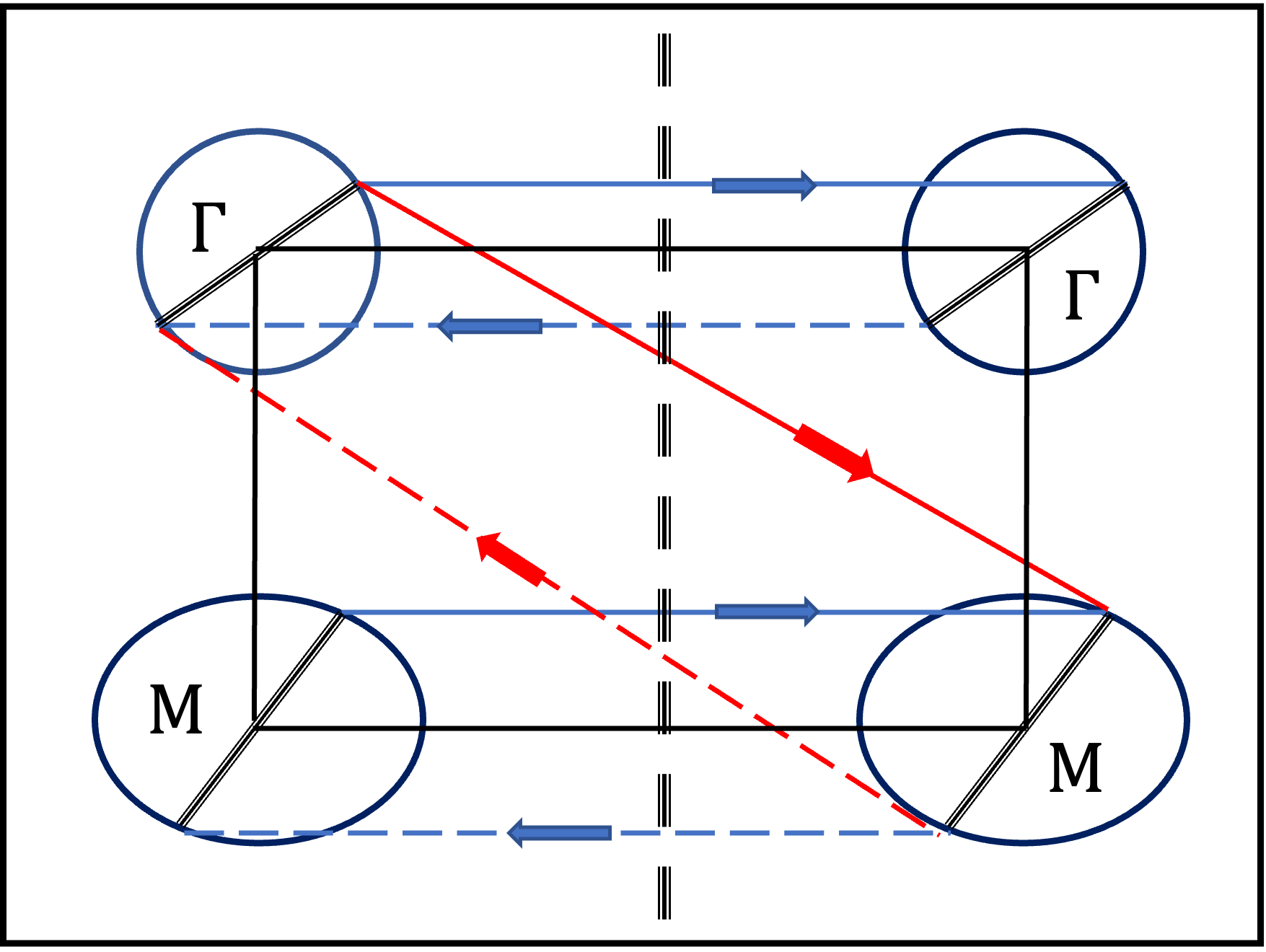}\\
    \caption{
    The schematic representation of the Andreev bound states (ABS) for the cases of (a) pure intra-pocket and (b) pure inter- pocket transmissions. The solid arrows correspond to electrons and dashed arrows to holes. For pure intra-pocket transmission (a) the ABS loop consists of electron and hole modes from same pocket (either $\Gamma$ or $M$ pocket). In contrast for pure inter-pocket transmission (b) electron and hole modes on both sides of the DW are entirely of the opposite pockets.}\label{fig:elhpockets}
\end{figure}

In this section we build the intuitive mechanism that can give rise to a $\pi$-junction at the DW by contrasting the limiting cases of scattering. For simplicity of our analytical description we consider the case where there is one $\Gamma$ pocket and one $M$ pocket at the Fermi level with sign-changing $s$-wave OP between them. In such a scenario let us consider the two limiting cases of DW induced scattering, viz., i) pure intra-pocket transmission and ii) pure inter-pocket transmission as shown in Fig.~\ref{fig:elhpockets}. The former scenario (i.e. (i)) is equivalent to the case of domains or defects that can cause an electronic phase scattering phase shift in $s$-wave superconductors where there is no sign change in order parameters. We will find that the $0$-junction is always favored in this case.

To determine the phase difference between the SC gaps across the DW, we analytically estimate the ABS energies in a setting of $s_{\pm}$ pairing and argue that in case of pure inter-pocket transmission, the energy of the occupied ABS levels is minimized when the overall phase-shift in OP is $\pi$ across the DW, as opposed to 0-phase shift that is stabilized in case of intra-pocket transmission. To show this, we use the well-known result  \cite{Beenakker1991} that the energy of a short JJ is dominated by the energy of the ABSs in the junction. This simplifies the calculation to the analysis of the bound states and allows us to ignore the extended states in the superconducting regions.

The ABSs, for the cases of pure intra- and inter-pocket transmissions, can be represented as effectively single-channel loops consisting of electron and hole modes as shown in Fig. \ref{fig:elhpockets} as blue and red loops. The solid line in a loop is the electronic mode with the arrow pointing along its direction of propagation whereas the dashed line is its time-reversed hole counterpart propagating in the opposite direction. The electron and hole are connected by Andreev reflection by the SC gap at the normal - superconductor interface. In the Fig. \ref{fig:elhpockets} the vertical split line in the middle denotes the domain-wall where the Josephson junction is formed. The circular contour schematically represents the hole-like $\Gamma$ pocket and the elliptical contour, the electron-like $M$ pocket on both sides of the DW. The blue loops represent the ABSs formed when the DW does not mix the modes between the $\Gamma$ and $M$ pocket at all during transmission. On the other hand the red loop represents the case when the DW completely converts a $\Gamma$ - pocket mode to an $M$ - pocket mode and vice-versa. Hence, the blue loops are relevant for analyzing pure intra-pocket transmission and the red loop for pure inter-pocket transmission. In both these case we can approximately determine the energies of the ABSs as we discuss below.

If we assume that the magnitudes of the SC gaps for the two pockets are same on both sides, the dependence of the ABS energy on the junction phase-shift for single channel loops as shown in Fig.~\ref{fig:elhpockets} is given by the single channel formula, $|\epsilon_{\text{ABS}} |=\Delta [1 - |t|^2 \sin ^2 \phi /2]^{1/2} $  \cite{Beenakker1991} where $|t|$ is the corresponding transmission amplitude through the junction, $\Delta$ is the SC gap and $\phi$ is the phase-twist seen by the electrons in the ABS loop.
This means that ABS that are formed by normal reflection (i.e. with $|t|\sim 0$) from the DW are not sensitive to the phase difference across the DW and can be ignored when determining the phase configuration. 

 For an $s_\pm$ order parameter the OP at the $M$ pocket has a $\pi$ phase shift relative to the $\Gamma$ pocket. This is in addition to any superconducting phase difference across the DW in Fig.~\ref{fig:elhpockets}. 
Since the Andreev reflection in Fig.~\ref{fig:elhpockets} picks up this phase associated with the superconducting OP, the phase $\phi$ for the ABS generated by each loop in Fig. \ref{fig:elhpockets} contains an additional phase (i.e. $\phi\rightarrow\phi+\pi$) for only the inter-pocket Andreev reflection processes (represented by the red loop) relative to the intra-pocket phase difference (shown by blue loops in Fig. 2). In addition, the Andreev process also can include a phase shift associated with the momentum mismatch of the $\Gamma$ and $M$ pocket together with the half-lattice shift across the DW. However, the time-reversal invariance of the Cooper pair means that the electron and holes originate from states with opposite momenta in the Brillouin zone. This results in a cancellation for any normal scattering phase accumulated in the Andreev reflection process. which manifests in the $|t|$ dependence of the ABS energy. This is consistent with the fact that the $0$-phase is the favored energy minimum state for intra-pocket (or $s$-wave) ABSs. 

The energy for the intra-pocket ABSs shown in Fig. 2 is determined by the phase difference $\phi$ across the DW, which is then minimized when $\phi=0$, i.e., a $0$-junction when this is the dominant process. In contrast, the dominance of inter-pocket ABSs leads to ABS energies with an effective phase difference $\phi+\pi$. This leads to the energy being minimized when the phase is $\phi=\pi$ leading to a $\pi$- junction when inter-pocket transmission is dominant.

The understanding of these two limiting cases of scattering motivates our proposed mechanism that given the $s_{\pm}$ pairing state between the two Fermi pockets in FTS, the DW-induced inter-pocket transmission effectively couples the OPs of opposite sign via Andreev reflection of inter-pocket transmitted modes and tends to favor a $\pi$-junction over a trivial junction with no phase difference. The discussion in this section is based on an ideal model that ignores several factors such as deviations from pure inter-pocket or intra-pocket transmission, difference in superconducting gap between pockets, dispersion of the ABS energy with momentum along the junction as well as the large superconducting gap size relative to Fermi energy. In addition, whether $0$- or $\pi$- junctions actually dominate in FTS depends on the 
intra-pocket versus inter-pocket scattering.
In Sec.~\ref{sec:numerical_results}, we compute the ABS spectrum by numerically solving the BdG equations with our effective model of the half-shift DW.
Apart from a qualitative validation of the results of this section, we will find the $\pi$- junction to be stable over a large parameter range for several models of FTS Fermi surface. 

In the following section, we describe the effective model of an FTS monolayer that we use for the BdG calculation.

\begin{figure}[t]
    \centering
    \includegraphics[width = \columnwidth]{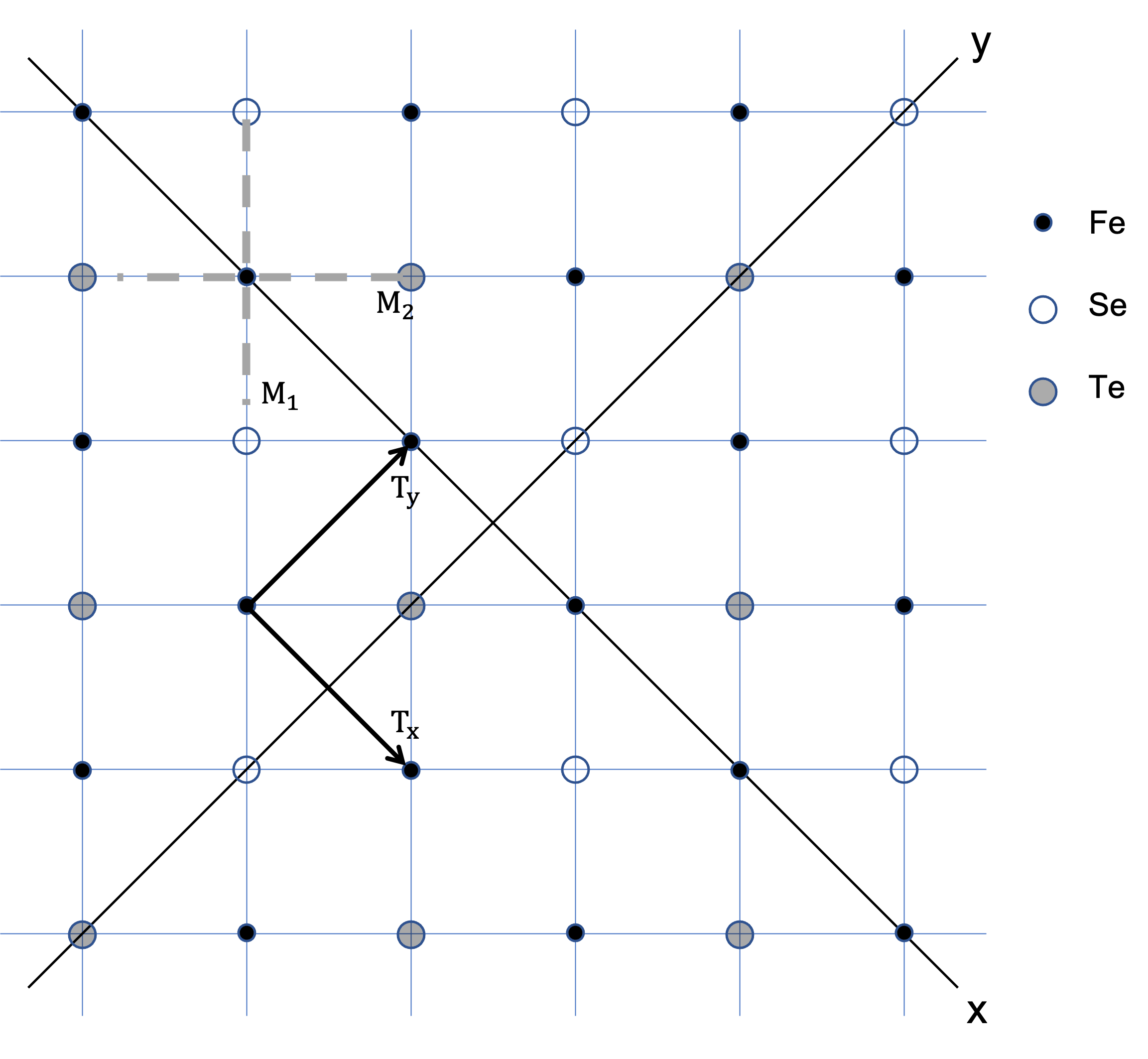}
    \caption{The lattice structure of an FTS monolayer. Fe-atoms are marked by dark circles, Se/Te atoms that are located off the Fe-plane alternately above and below the plane,by the shaded circles for above and hollow circles for below. The half-translations $T_{x/y}$ and the mirror planes $M_{1,2}$ are also shown.}
    \label{fig:FTS_monolayer_lattice}
\end{figure}

\begin{figure}[t]
	\centering
 \includegraphics[width = \columnwidth]{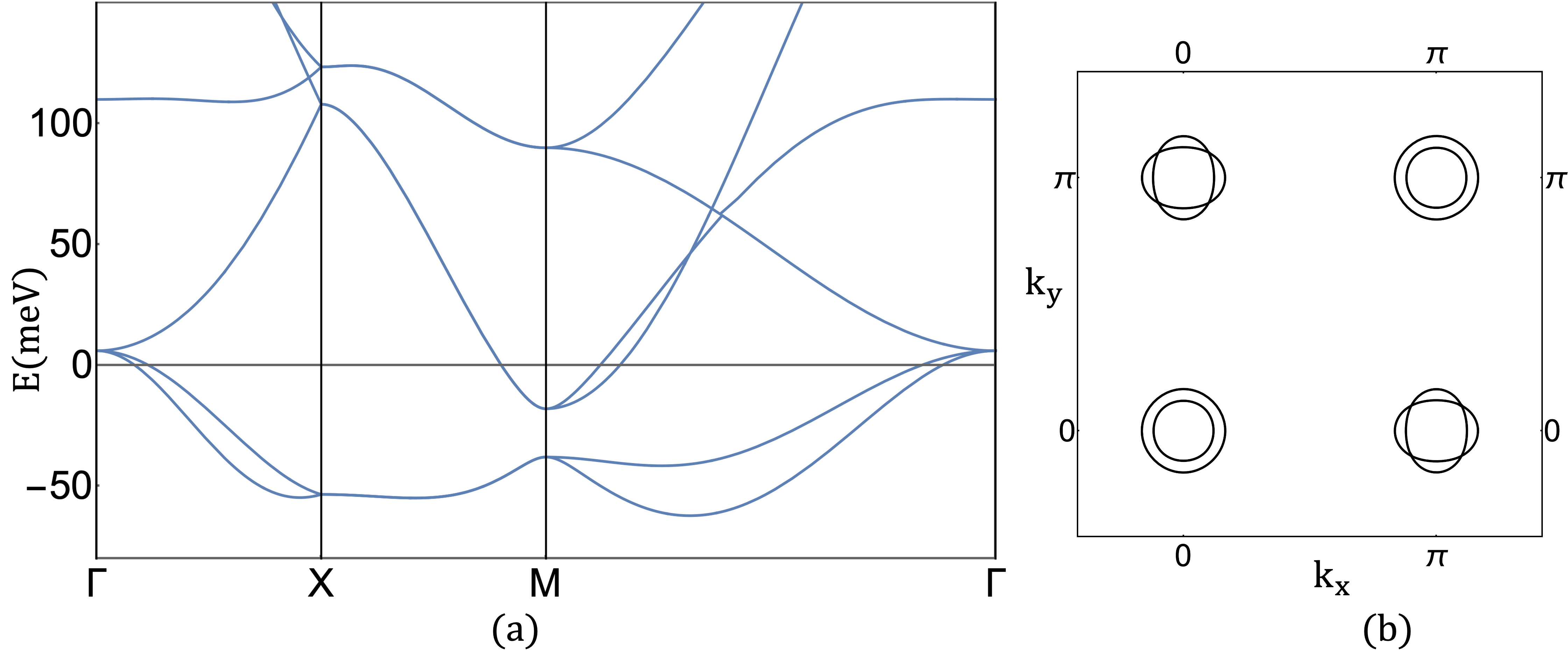}

    \caption{The band structure of FTS monolayer along the high symmetry line $\Gamma X M \Gamma$ showing the two hole-like pockets around $\Gamma$ point and two electron-like pockets around $M$ point (a). The Fermi contour is shown where the $\Gamma$ pockets are circular and $M$ pockets are slightly elliptical (b).}\label{fig:Bands_Fermi_pockets}
\end{figure}

\section{Microscopic model for the HUCS DW\label{sec:effective_model}}

In this section we discuss the construction of our effective model for the band structure around the Fermi level of the HUCS DW in an  FTS monolayer. This model is used for simulating the effects of the half-shift DW. Due to the negligible dispersion of the FTS bands perpendicular to the Fe-plane, the Fermi sheets of 3D FTS are approximately cylindrical. Hence, an effective model can be simplified by reducing the problem to two-dimensional FTS monolayer without sacrificing any relevant features of the Fermi pockets. With that simplification we construct a tight binding model that is determined by the symmetries of the system.

While choosing the basis for our model we rely on previous first-principle analysis of orbital content \cite{PRL_LDA_FTS,FTS_LDA_PRB} which reveals that near the Fermi level the bands dominantly contain the Fe-$3d_{x^2 - y^2}$ and $3d_{xz/yz}$ orbitals. This claim is also supported by polarization selective ARPES measurements of the band intensity near the high symmetry points \cite{PRL_LDA_FTS}. The lattice structure of a two-dimensional monolayer is shown in Fig. \ref{fig:FTS_monolayer_lattice}.

The spatial symmetries of the FTS monolayer as pointed out in Fig. \ref{fig:FTS_monolayer_lattice} are \cite{FTS_monolayer} i) the 3D inversion, ii) two fold rotations around x and y axes that pass through the Fe atomic sites (see Fig. \ref{fig:FTS_monolayer_lattice}), iii) the mirror reflections across the next-nearest neighbor Fe-Fe bond axes, $\hat{M}_{1,2}$ (Fig. \ref{fig:FTS_monolayer_lattice}) and iv) the glide-reflections ($\hat{T}_{x/y}\hat{M}_z$). We note that the glide-reflections can be thought of as a combination of a half-translation along x or y-axis ($\hat{T}_{x/y}$ in Fig. \ref{fig:FTS_monolayer_lattice}) and a mirror reflection across the xy-plane ($\hat{M}_z$). Due to the alternating up and down out-of-plane atomic positions of the chalcogens (Fig. \ref{fig:FTS_monolayer_lattice}) the half translations by themselves are not exact symmetries of the system. The asymmetry after the half-translation gives rise to a sublattice structure to our Fe-site based lattice model. In fact the HUCS defect can be characterized by one side of the lattice transformed by a half translation.

To formalize the model Hamiltonian we denote the basis Fe-orbitals by $\phi ^{\alpha}_i$ where $\alpha \, (= a, \, b)$ denotes the sub-lattice index of the Fe-sites and $i \, (\in \{1,2,3\})$ denotes the orbital index such that $\phi _1 \equiv 3d_{x^2 - y^2} $, $\phi _{2/3} \equiv 3d_{xz/yz} $. The matrix elements of the Hamiltonian, $\langle \phi ^{a/b}_i| \hat{H}|\phi ^{a/b}_j \rangle $, are determined by the symmetry constraints of the system that are listed above. The construction of matrix elements proceeds by considering all hopping amplitudes between neighboring orbitals and imposing any constraints due to the symmetry properties of the Hamiltonian. We only consider hoppings along nearest neighbor (NN) bonds and next-nearest neighbor (NNN) bonds. We do not explicitly derive the matrix elements here as the same steps for the case of FeSe, which has same symmetries as FTS, have been described in detail in \cite{FTS_monolayer}. 

Considering the nearest neighbor (NN) and next-nearest neighbor (NNN) hoppings between all pairs of orbitals from each sub-lattice, the resultant momentum-space Hamiltonian matrix elements, $H^{\alpha \beta}_{ij}(k) \equiv \langle \phi _i^{\alpha} (k)|\hat{H}(k)|\phi _j^{\beta}(k) \rangle $, have the following form,
\begin{align}\label{eq:monolayer_Ham}
    \begin{split}
        & \text{Intra-sub-lattice hoping terms:} \\
        & H^{aa}_{ii} = H^{bb}_{ii} = \epsilon _i + 4 u_i \cos k_x \, \cos k_y \quad : \quad i = 1,2,3\\
        & H^{aa}_{12} = 4iu \cos k_x \sin k_y = -H^{aa}_{21} = -H^{bb}_{12} \\
        & H^{aa}_{13} = 4iu \sin k_x \cos k_y = -H^{aa}_{31} = -H^{bb}_{13} \\
        & H^{aa}_{23} = -4u_{23} \sin k_x \sin k_y = H^{aa}_{32} = H^{bb}_{23} \\
        & \text{Inter-sub-lattice hoping terms:} \\
        & H^{ab}_{11} = 2 t_1 (\cos k_x + \cos k_y) \\
        & H^{ab}_{22,33} = 2 (t_{x,y} \cos k_x + t_{y,x} \cos k_y) \\
        & H^{ab}_{12} = H^{ab}_{21} = -2i t \sin k_y \\
        & H^{ab}_{13} = H^{ab}_{31} = 2i t \sin k_x
    \end{split}
\end{align}
Due to rotational symmetries we have $\epsilon _2 = \epsilon _3 $ and $u_2 = u_3 $. Furthermore, among all the matrix elements listed above $H^{aa}_{12,13}, H^{ab}_{12,13}$ are matrix elements of $\hat{H}_-$ whereas the rest are of $\hat{H}_+$ \cite{FTS_monolayer}. 

Here we note a useful decomposition of the Hamiltonian depending on its behaviour under half-translation operators, $\hat{T}_{x/y} $. Since half-translations applied twice (i.e., $\hat{T}^2_{x/y}$) are exact symmetries of the Hamiltonian,  $\hat{H} $ can be decomposed into two parts:
\begin{align}\label{eq:lat_Ham_decomp}
    \hat{H} =\hat{H}_+ + \hat{H}_-
\end{align}
such that $\hat{H}_+ $ ($\hat{H}_-$) is even (odd) under $\hat{T}_{x/y}$. This distinction implies that to simulate the HUCS DW, the terms corresponding to $\hat{H}_- $ should flip sign compared to pure Hamiltonian across the DW whereas the $\hat{H}_+ $ would remain the same as the defect-free case. This is because if $\hat{H}_D $ denotes the defect Hamiltonian modelling the HUCS DW along the $\hat{y}$-axis, it is related to the defect-free Hamiltonian as $\hat{H}_D(x>0) = \hat{T}_y\hat{H}(x>0)\hat{T}_y^{-1} $ and thus $\hat{H}_-$ would flip sign on one side of the DW.

After the construction of our Hamiltonian matrix, we tune the model parameters that produces a band structure reasonably simulating the key features of ARPES and DFT based results, viz. i) the relative Fermi energies of the electron-like and hole-like bands and ii) the shape and size of the Fermi pockets. The band structure resulting from our model is shown in Fig. \ref{fig:Bands_Fermi_pockets}. Our model simulates total six bands originating from three orbitals and two sublattices, four of them form the two hole-like pockets around $\Gamma$ and two electron-like pockets around $M$. The Fermi energy of the hole-like pockets are about 3 fold smaller than that of the electron-like pockets. The $\Gamma $ pockets are circular in shape and the $M$ pockets are slightly oblate - both the $\Gamma $ and $M$ pockets cover same area which is consistent with ARPES estimation \cite{PRL_LDA_FTS}. Using this simulated model we can perform required calculation of the DW-induced scattering and the BdG spectrum for junctions.

\section{Numerical results\label{sec:numerical_results}}

\subsection{Solving the scattering problem\label{sec:scattering_problem}}

To treat the DW-induced scattering problem the Hamiltonian is discretized perpendicular to the DW and periodic boundary condition can be imposed along the direction of the DW. We use the Python based free KWANT package \cite{KWANT2014} to numerically solve the scattering problem. In KWANT the crystalline half-shift DW (say the DW is along y-direction, thus, fixing $k_y$ for the Hamiltonian) can be modelled as two semi infinite leads (along x-axis) that are half-translated w.r.t. each other and joined together with no separate scattering region in between. The resultant scattering matrices are obtained as a function of $k_y $ and energy, $E$. In the limit where the SC gap is much smaller than the Fermi energies, the scattering matrix only at the Fermi level (i.e., $E=0$) is sufficient for our analysis as the ABS modes in that case consist of modes only at the Fermi level. Corresponding to the four bands that cross the Fermi level, the scattering solution has a $4 \times 4$ transmission matrix at each energy slice. Denoting the momentum modes of $\Gamma$ pockets by $k_i$ and $M$ pockets by $q_j $ where $i,j \in \{1,2\} $ representing two pockets, the transmission amplitudes is denoted by $t_{k_i q_j}(k_y,E)$. To set a measure for the relative inter-pocket transmission strength we define the total intra-pocket and inter-pocket transmission rates as
\begin{align}
    \begin{split}
        & P_{\text{intra}}(E=0) = \sum _{k_y, i,j} [|t_{k_ik_j}(k_y,0)|^2 + |t_{q_iq_j}(k_y,0)|^2 ] \\
        & P_{\text{inter}}(E=0) = \sum _{k_y, i,j} [|t_{k_iq_j}(k_y,0)|^2 + |t_{q_ik_j}(k_y,0)|^2]
    \end{split}
\end{align}
In the extreme cases of pure intra-pocket and pure inter-pocket transmissions these quantities are $P_{\text{intra}} = 1, \, P_{\text{inter}} = 0$ and $P_{\text{inter}} = 1, \, P_{\text{intra}} = 0$ respectively. As we explain in Sec. \ref{sec:correlation} the relative difference between the two quantities, i.e., $\delta P = P_{\text{inter}} - P_{\text{intra}}$ can be used to quantitatively correlate to the junction stability.

\subsection{Calculation of the BdG spectrum\label{sec:BdG_spectrum}}
Now we describe the calculation of the junction BdG spectrum due to the DW. For this purpose we first incorporate an $s_{\pm}$ order parameter in our model. We choose the simplest form of pairing function, specified by $\Delta (k) = \Delta _0 + 2 \Delta _1 \cos k_x \cos k_y $, that flips sign between the $\Gamma $ and $M$ pockets for appropriately chosen value for $2|\Delta _1|/|\Delta_0| $. With a trivial structure in the orbital and sub-lattice space proportional to the identity matrix, the pairing function can be written as
\begin{align}\label{eq:spm_gap}
    \begin{split}
        & \Delta ^{ab}_{ij}(k_x, k_y) = \delta ^{ab} \delta _{ij} (\Delta _0 + 2 \Delta _1 \cos k_x \cos k_y)
    \end{split}
\end{align}

This spin-singlet pairing potential can be combined with $\hat{H}$ from Eq.~\ref{eq:lat_Ham_decomp}, the normal state Hamiltonian for FTS monolayer described in the last section, to construct a BdG Hamiltonian 
\begin{align}\label{eq:BdG_Ham}
&\hat{H}_{\text{BdG}}=(\hat{H}-\mu )\tau^z+\Delta \tau^x,
\end{align}
where $\tau^{x,z}$ are the Pauli matrices describing the particle-hole degree of freedom in Nambu spinor space \cite{schrieffer_SC}.

We use this BdG Hamiltonian in Eq. \ref{eq:BdG_Ham} along with the form for $\Delta$ given by Eq. \ref{eq:spm_gap} for the purpose of numerical evaluation of the BdG spectrum. As described before in our finite size modeling of the Josephson junction, the domain-wall is incorporated by shifting the lattice Hamiltonian (Eq. \ref{eq:monolayer_Ham}) by a half-translation operator across the intended defect location. Then, the phase difference between the SC OP on both sides of the domain-wall is set as $\Delta (x>0) = \Delta (x<0) \, e^{i\phi}$ and choose  $\phi = 0$ or $\phi = \pi$ to simulate the i) $0$-junction and ii) $\pi$-junction respectively.

\begin{figure}
    \centering
    \includegraphics[width = \columnwidth]{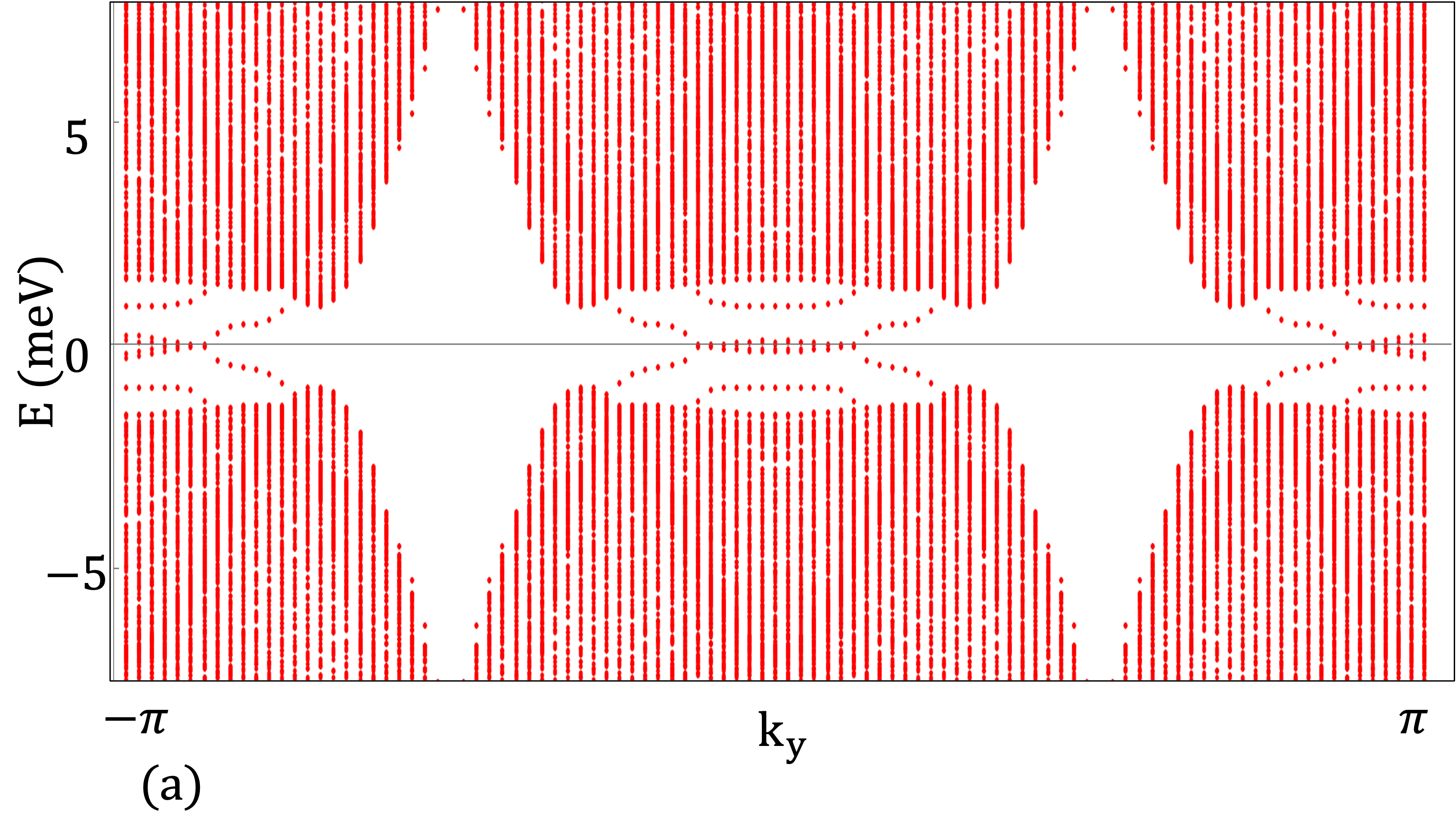} 
    \vfill
    \includegraphics[width =  \columnwidth]{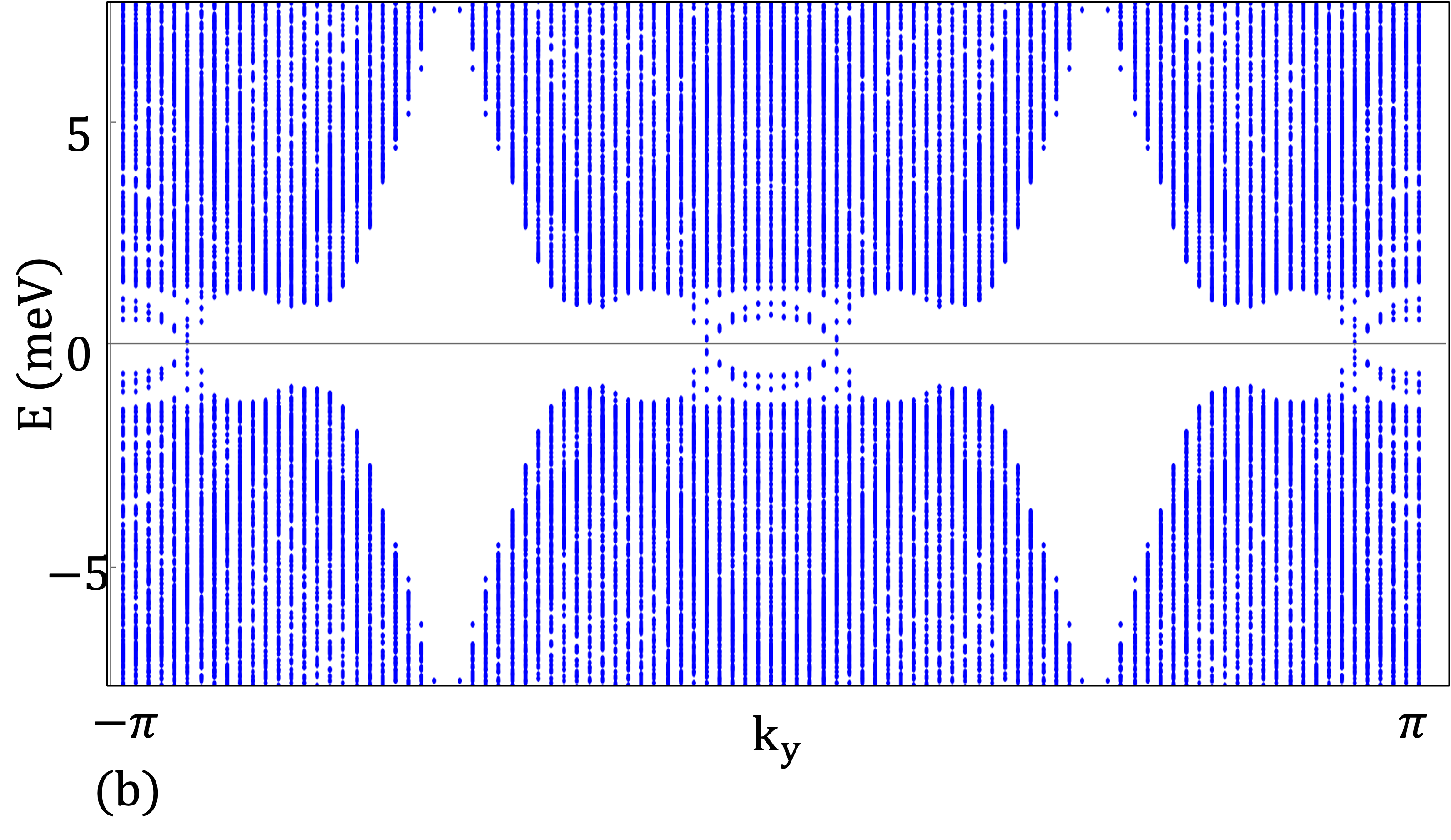}
    \caption{Typical BdG spectra for the cases of (a) $0$-junction and (b) $\pi$-junction. The sub gap energy levels are typically much closer for the case of 0-junction (a) than that of $\pi $-junction (b).}
    \label{fig:BdG_spectra_0_pi}
\end{figure}

The resultant BdG spectrum for a typical set of model parameters is shown in Fig. \ref{fig:BdG_spectra_0_pi} where the sub gap energy levels that appear around the Fermi level correspond to the Andreev bound states. The spectrum with red dots (Fig. \ref{fig:BdG_spectra_0_pi}(a)) corresponds to 0-junction whereas the one with blue corresponds to $\pi$-junction. A careful observation of the two spectrum reveals that the sub gap levels for the 0-junction, on an average, are much closer to the Fermi level than the ones for the $\pi$-junction which implies that the occupied ABS would have lesser energy for the case of $\pi$ junction favoring it to be energetically stable compared to the trivial junction with no phase shift. As mentioned earlier, the energies of the extended states (i.e., the levels outside the gap) are insensitive to the phase twist of the junction~\cite{Beenakker1991}. In our numerically obtained results as well, we find that the difference in energies (between $0$- and $\pi$-junctions) due to the bulk modes are negligible compared to the ABS modes.

Thus, to formalize the junction stability the difference of the occupied ABS energy levels between the 0- and $\pi$-junction, is a sufficient indicator of the junction stability. The difference is given by
\begin{align}
    \begin{split}
        & \delta E \approx E^{\pi}_{\text{ABS}} - E^{0}_{\text{ABS}}=\sum_{k_y,n} [|E^n_{\text{ABS}}(k_y, \pi)| - |E^n_{\text{ABS}}(k_y,0)|] 
    \end{split}
\end{align}

where $n$ spans the discrete labels denoting the ABS levels. The sign of $\delta E$ would determine which junction has lesser energy as $\delta E<0 $ and $\delta E > 0 $ would imply $0$- junction and $\pi$-junction are preferred respectively. 
For the model parameters that simulate a band structure as shown in Fig. \ref{fig:Bands_Fermi_pockets}, $\delta E >0 $ implying that $\pi $-junction is thermodynamically stable. 

\subsection{Analysis of the numerical results\label{sec:correlation}}

\begin{figure}
\vspace{.1 in}
    \centering
    \includegraphics[width = \columnwidth]{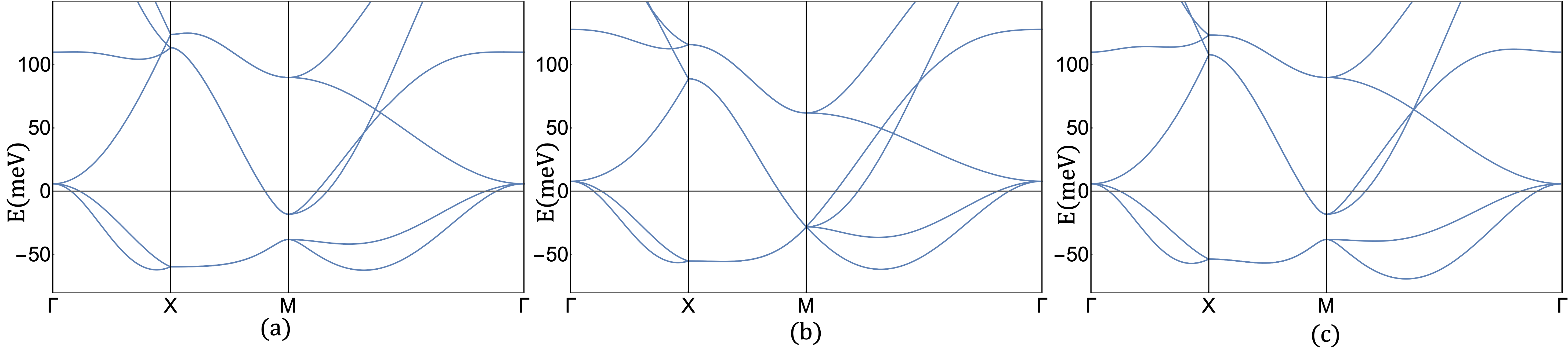} \\
    \includegraphics[width = \columnwidth]{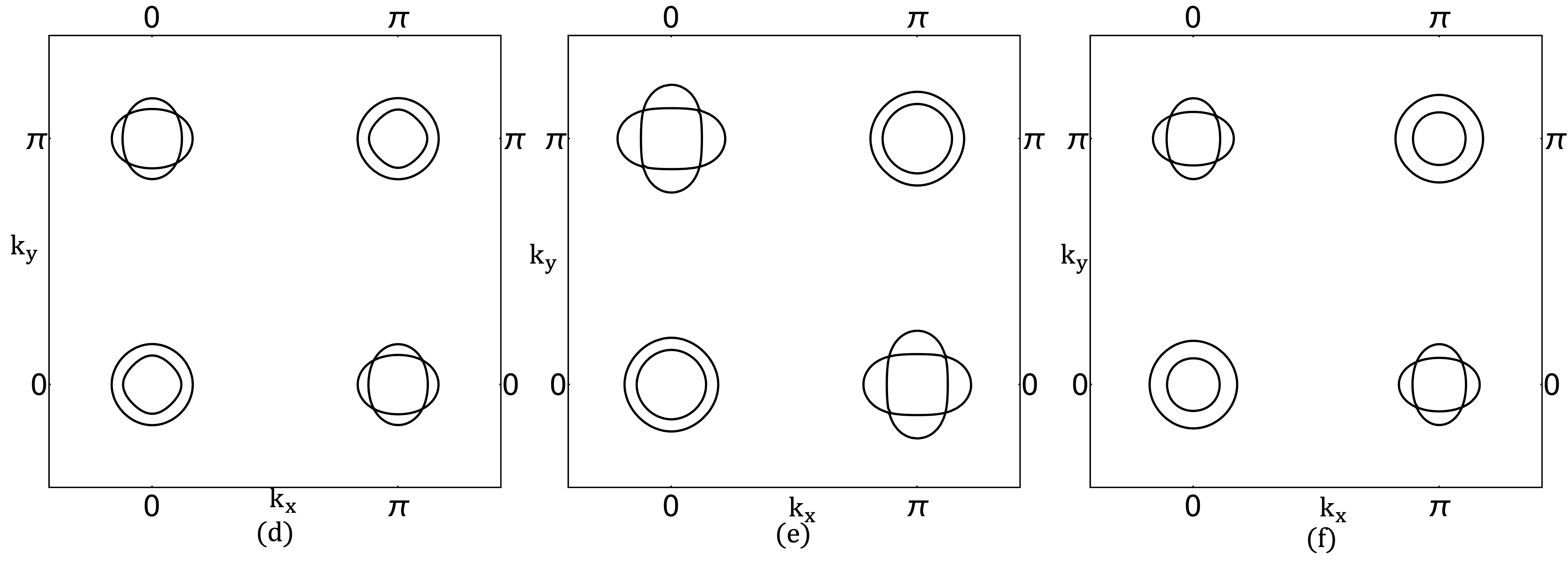} 

    \caption{A sample of resultant band structures (a)-(c) and the corresponding Fermi contours (d)-(f) due to variation of the model parameters.}
    \label{fig:parameter_variation}
\end{figure}

 While the results in the previous sub-section show that a $\pi$-junction might be the favored state for an HUCS DW for certain parameters, the favorability of the $\pi$-junction relative to a $0$-junction is non-universal and depends on parameters of the system. 
 To estimate the likelihood of whether a $\pi$-junction is realized in the FTS experiment~\cite{Madhavan2020} we vary our model parameters around the simulated band structure and repeat the BdG spectrum analysis for each case.  As mentioned in the last section the model parameters were varied keeping the key observable features such as the relative Fermi pocket sizes between the electron and hole pockets and the relative Fermi energies of the bands consistent with ARPES so that each set of parameters still reasonably simulate the FTS monolayer. The parameter variations in this case are mainly attributed to the changes in band curvatures, the shapes of the Fermi pockets and orbital projections of different bands.  Such changes can alter the DOS around the Fermi pockets as well as the DW-induced coupling between states of various pockets thus affecting various scattering strengths. Thus, even within the FTS monolayer regime such variations can alter the scattering outcomes to a certain extent.
 While varying the parameters we constrained the resultant Fermi pocket sizes and the relative Fermi energies of the hole-like and electron-like bands to be consistent with the ARPES data. Thus, the model variations amount to changing other degrees of freedom such as the band curvatures, Fermi pocket shapes, orbital projections on different bands, etc. Few samples of the band structure are shown in Fig. \ref{fig:parameter_variation}.

\begin{figure}
\vspace{.1 in}
    \centering
    \includegraphics[width = .9 \columnwidth]{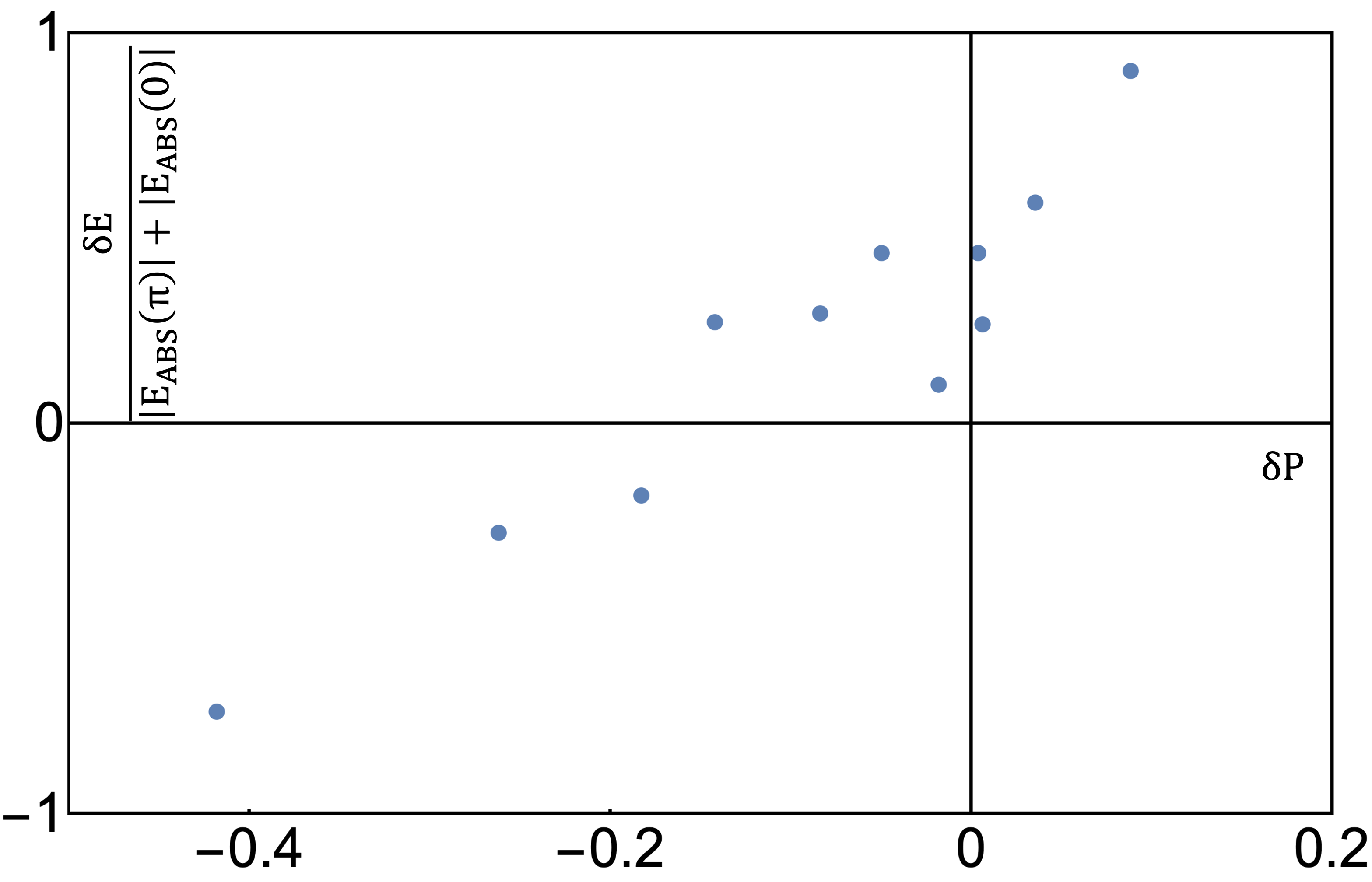}
    \caption{The difference in occupied energy of $\pi$-junction relative to $0$-junction, scaled by the total energy of the ABSs for the two junctions, $\delta E / [|E_{\text{ABS}}(\pi)| + |E_{\text{ABS}}(0)|]$, is plotted along y-axis and the corresponding inter-pocket transmission strength relative to the intra-pocket transmission, $\delta P = P_{\text{inter}} - P_{\text{intra}}$, is plotted along the x-axis.}
    \label{fig:correlation}
\end{figure}

The results from the BdG spectrum calculation (shown in Fig.~\ref{fig:correlation}) reveal that there is a large fraction of parameter sets for which $\delta E>0$ (i.e., $\pi$-junction is favored over 0-junction) which appears in Fig. \ref{fig:parameter_variation} as the points below the x-axis. This implies that $\pi$-junction is robust to the model variations considered here. Furthermore, when the values for $\delta E $ are plotted along with the corresponding values of $\delta P$ we find a strong correlation between the inter-pocket transmission strength with the $\pi$-junction stability. This quantitative correlation supports the fact that the inter-pocket mixing via the domain-wall tends to favor the $\pi$-junction which is consistent with the our model for the $\pi$-shift described in Sec.~\ref{sec:effective_model}.

\section{Summary and Conclusion}
In this paper we have proposed a mechanism favoring $\pi$-shift in the order parameter (OP) in an FTS monolayer across a HUCS DW of the lattice. The mechanism works in the background of $s_{\pm}$ pairing state when the s-wave OPs of the two pockets have opposite signs. We argued that if the DW can mix the modes of the two pockets during transmission strongly enough, the two opposite signed OPs on both sides get coupled as a consequence of Andreev reflection. In the case when the opposite signed OPs are mixed strongly across the DW, an overall phase shift of $\pi$ stabilizes the energies of the Andreev bound states (ABS) associated with the junction. We numerically justified this mechanism by modelling an FTS monolayer and computing the BdG spectrum along with the scattering strengths due to the DW. By varying our model parameters within the range of FTS monolayer we found that majority of the parameters support $\pi$-junction manifested as the ABS levels for the $\pi$-junction having lesser energy than that of $0$-junction. Furthermore, by comparing with the inter-pocket transmission strength w.r.t. intra-pocket transmission we found that the strong inter-pocket transmission tend to favor the $\pi$-junction which supports our mechanism as described above. It should however be noted that this conclusion is non-universal and depends on parameter choice. While we find the $\pi$-junction to be stable for most parameter choices, there are also parameters where the $0$-junction is favored.

Our results provide a possible quantitative model for the bright features measured~\cite{Madhavan2020} in HUCS DWs, which has been interpreted as a $\pi$-junction on the surface of a TSC. Such a $\pi$-junction on the surface of a TSC is predicted to have a gapless helical mode~\cite{Fu_Kane_theory} that may play a key role in topological quantum information processing. Our calculation focuses on the phase difference across a DW for a single layer FTS model. Existence of a $\pi$-junction at a half-shift domain wall in a monolayer of FTS is thus the primary result of our work. Even though we have ignored the inter-layer coupling that is crucial for proximity-induced superconductivity on the topological surface state of FTS \cite{Dongfeietal2018}, the inter-layer coupling is 
weak compared to the in-plane tunneling and hence, we expect the $\pi$-junction to continue to be favored for adjacent monolayers with the DW defect interface. Presumably the DW defect 
would extend a finite depth into the bulk of the material, beyond which the surface state would heal. More detailed superconducting phase sensitive measurements such as corner junction devices or high resolution scanning tunneling spectroscopy would be able to distinguish this topological mechanism from competition with a strain-induced nematic state \cite{Nematic_transition}. Another interesting direction would be to study how the $\pi$-shift in the bulk SC OP is proximity-induced on the TI surface state in FTS. The inter-play of the distinct DOS for ABSs from the FTS model shown in Fig.~\ref{fig:BdG_spectra_0_pi} in addition to the helical Majorana states arising from the TI surface state.

\begin{acknowledgments}
This work was supported by NSF QII-TAQS 1936246. JDS would also like to acknowledge NSF DMR-1555135 (CAREER) for support. We also thank  Ruixing Zhang, Jennifer Hoffman, Benjamin November and Jinying Wang for valuable discussions.
\end{acknowledgments}

\bigskip
\bibliography{main}

\end{document}